\documentclass{ifacconf}
\usepackage{array}
\newcommand{\PreserveBackslash}[1]{\let\temp=\\#1\let\\=\temp}
\newcolumntype{C}[1]{>{\PreserveBackslash\centering}p{#1}}
\newcolumntype{R}[1]{>{\PreserveBackslash\raggedleft}p{#1}}
\newcolumntype{L}[1]{>{\PreserveBackslash\raggedright}p{#1}}
\usepackage{graphicx}
\newcounter{para}
\renewcommand{\thepara}{\roman{para}}
\newcommand\mypara{\par\refstepcounter{para}\thepara.\space}

\usepackage{ragged2e}
\usepackage{float}

\usepackage{ragged2e}
\usepackage{float}
\graphicspath{{./fig/}}
\usepackage{tabularx}
\usepackage{graphics}
\usepackage{parskip}
\usepackage{rotating} 
\usepackage{makecell}
\usepackage{enumerate}
\usepackage{longtable}
\usepackage{lettrine}
\usepackage{multirow}
\makeatletter
\def\thickhline{%
  \noalign{\ifnum0=`}\fi\hrule \@height \thickarrayrulewidth \futurelet
   \reserved@a\@xthickhline}
\def\@xthickhline{\ifx\reserved@a\thickhline
               \vskip\doublerulesep
               \vskip-\thickarrayrulewidth
             \fi
      \ifnum0=`{\fi}}
\makeatother

\newlength{\thickarrayrulewidth}
\usepackage{lipsum}
\usepackage{textcomp}

\makeatletter
\let\old@ssect\@ssect 
\makeatother

\usepackage{natbib}
\usepackage[dvipsnames]{xcolor}
\RequirePackage[bookmarks, bookmarksopen=true, plainpages=false, pdfpagelabels]{hyperref}
\hypersetup{
colorlinks=true,
linkcolor=blue,
filecolor=Blue,      
urlcolor=Black,
citecolor = red,
}

\makeatletter
\def\@ssect#1#2#3#4#5#6{%
  \NR@gettitle{#6}
  \old@ssect{#1}{#2}{#3}{#4}{#5}{#6}
}
\makeatother
\usepackage[hyphenbreaks]{breakurl}
\renewcommand\thesubsubsection{\thesubsection.\arabic{subsubsection}}

\let\oldciteauthor=\citeauthor
\def\citeauthor#1{\hypersetup{citecolor=black}\oldciteauthor{#1}}
\let\oldcite=\cite
\def\cite#1{\hypersetup{citecolor=black}\oldcite{#1}}

\def\BibTeX{{\rm B\kern-.05em{\sc i\kern-.025em b}\kern-.08em
    T\kern-.1667em\lower.7ex\hbox{E}\kern-.125emX}}

\usepackage[]{amsmath}
\usepackage{amssymb,amsfonts, mathtools,}
\usepackage{bbm}
\usepackage{bm}
\usepackage{cuted}
\usepackage{etoolbox}

\usepackage{tikz}
\usetikzlibrary{shapes,arrows,external,calc}
\usepackage{pgfplots}
\pgfplotsset{width=10cm,compat=1.9}
\tikzstyle{block} = [draw, fill=green!20, text centered, rectangle,
minimum height=0.8cm,minimum width=6em, align=right]
\usepgfplotslibrary{external} 
\usepackage{epsfig}

\usepackage{soul}

\begin{document} 
\begin{frontmatter}
\thanks[footnoteinfo]{This work was not funded by any organization.}
\thanks[authorinfo]{C. Enwerem is with the Department of Electrical and Computer Engineering, University of Maryland, College Park, MD 20742, USA. He was with the Department of Electrical Engineering, University of Nigeria, Nsukka, Enugu 410001, Nigeria when this work was conducted.}
\thanks[corrauth]{Corresponding author.}

\title{Optimal Controller Tuning Technique for a First-Order Process with Time Delay\thanksref{footnoteinfo}}

\author[First]{Clinton Enwerem\thanksref{authorinfo}}
\author[Second]{Ihechiluru Okoro\thanksref{corrauth}}

\address[First]{Department of Electrical and Computer Engineering, University of Maryland, College Park, MD 20742, USA (email: enwerem@umd.edu).}

\address[Second]{Department of Electrical Engineering, University of Nigeria, Nsukka, Enugu 410001, Nigeria (email: ihechiluru.okoro@unn.edu.ng).}

\thispagestyle{plain}
\pagestyle{plain}

\begin{abstract}
We present a controller tuning strategy for first-order plus time delay (FOPTD) processes, where the time delay in the model is approximated using the Padé function. Using Routh-Hurwitz stability analysis, we derive the gain that gives rise to desirable PID controller settings. The resulting PID controller, now correctly tuned, produces satisfactory closed-loop behavior and stabilizes the first-order plant. Our proposed technique eliminates the dead-time component in the model and results in a minimum-phase system with all of its poles and zeros in the left-half $s$-plane. To demonstrate the effectiveness of our approach, we present control simulation results from an in-depth performance comparison between our technique and other established model-based strategies used for the control of time-delayed systems. These results prove that, for the FOPTD model, Padé approximation eliminates the undesirable effects of the time delay and promises a faster tracking performance superior to conventional model-based controllers.
\end{abstract}
\end{frontmatter}

\section{Introduction}
In the process industry, the first-order plus time delay (FOPTD) representation is popular for plant modeling because it affords a truncation of higher-order dynamics and distribution of the same between a time constant and a delay term (\cite{ogataModernControlEngineering2015}). However, in the tuning of such time-delayed processes, traditional and early tuning methods, such as Ziegler-Nichols (Z-N), have led to performance complications and hence, cannot be applied directly to the FOPTD model structure, which is unstable both in the open-loop case and in feedback systems with proportional-only controllers (\cite{grimholtOptimalPIControlVerification2012, StepResponsePlot}). As a result, virtually most of the PID controllers used in these FOPTD-approximated process plants are poorly tuned owing to transport delays and recycle loops (\cite{grimholtOptimalPIControlVerification2012, wangIMCPIDController2016, medarametlaNovelPIDController2018}). There is, therefore, a need to formulate control tuning procedures that will handle the dead-time component effectively while guaranteeing stability and satisfactory control performance. To address these control requirements, we present two independent controller design scenarios – (i) when the time delay in the FOPTD process is approximated using Padé approximation, and (ii) when the delay is retained. The former is dependent on the Routh-Hurwitz (R-H) stability criterion, and the latter, on the gain margin and crossover frequency of the FOPTD-model's Nyquist plot. As we will show in this paper, approximating the time delay offers significant performance advantages over the controller-tuning scheme where the time delay is retained. 

Significant research efforts have been devoted to the PID control of FOPTD processes, however, an exhaustive review of these methods is outside the scope of this work. Interested readers may consult (\cite{belwalModelingControlFOPDT2023}) for a detailed survey on the topic, and (\cite{odwyerHandbookPIPID2009}), for an in-depth study on the subject of PID controller tuning in general. In (\cite{sharmaModelbasedApproachController2013}), the authors compare the performance of different Internal Model Control (IMC)-based tuning rules applied to regulate an FOPTD process. These rules are derived from the original IMC design proposed by (\cite{garciaInternalModelControl1982}). Here, the delay is not approximated, leading to a sluggish system response. A proportional-integral (PI) controller, tuned via Ziegler-Nichols law, is used to control an FOPTD model in (\cite{yuceFractionalOrderPI2016}). With the delay in the system, their approach leads to a slow and highly unstable output as opposed to a faster and reasonably-damped response obtained via approximation, as will be presented in this paper. The control of an FOPTD process is achieved via an IMC-PID controller in (\cite{wangIMCPIDController2016}). While the system is rendered stable with this method, the lack of approximation of the delay term takes a toll on its response speed. Similarly, in (\cite{tavakoliOptimalTuningPID2003}), dimensional analysis is utilized to obtain optimal PID controller settings with delay term retained in the FOPTD model, but the system response is observed to be slow. More recently, a PID controller cascaded with a second-order filter is proposed in (\cite{medarametlaNovelPIDController2018}) for stable and unstable FOPTD processes. The controller stabilizes the plant in the unstable case, but at the expense of the system response speed. More recently, in (\cite{maPIDControlDesign2022}), the authors present a multiplicity-induced dominancy controller tuning technique for neutral delay systems, with delay robustness and stability considerations.

In this paper, we present an effective strategy for feedback control of first-order systems with time delay, which results in a simple PID controller that is expertly tuned, produces satisfactory closed-loop behavior, and stabilizes the plant, while eliminating the undesirable impeding effects of dead time. In particular, we consider retarded delay systems (i.e., delay systems with a single delay and finitely-many right-half plane roots (\cite{maPIDControlDesign2022, kharitonovTimeDelaySystemsLyapunov2012}) as opposed to the neutral delay (infinite) case. Our approach involves approximating the time delay via the Padé function and then deriving the controller gain using information from the Routh-Hurwitz stability criterion. To portray the method's effectiveness, we present results from an exhaustive comparison between the system response characteristics with the dead-time-approximation technique and the response with other model-based approaches, namely Internal Model Control (IMC), following ideas presented in previous work (\cite{okoroInternalModelControl2019}), and Skogestad-Internal Model Control (SIMC). 
\subsection{Contributions}

Our main contributions are as follows: 
\begin{enumerate}[a.]
    \item Development of a controller tuning scheme based on Padé approximation and ideas from frequency-domain stability theory.
    \item Comparison of the proposed delay-approximation technique with IMC and SIMC approaches.
    \item Demonstration of the advantages of the approximation approach over conventional methods, using original numerical simulations. 
\end{enumerate}

The rest of this paper is organized thus. Modeling and controller tuning techniques
for the FOPTD process are developed in Section \ref{sec:foptdmod}. In Section \ref{sec:simex}, we present simulation experiments with results, analyses, and inferences. Finally, concluding notes and directions for further research are given in Section \ref{sec:conc}.

\section{FOPTD Modeling and Controller Tuning Methods}
\label{sec:foptdmod}

\begin{figure}[H]
    \centering
    \includegraphics[width=\columnwidth]{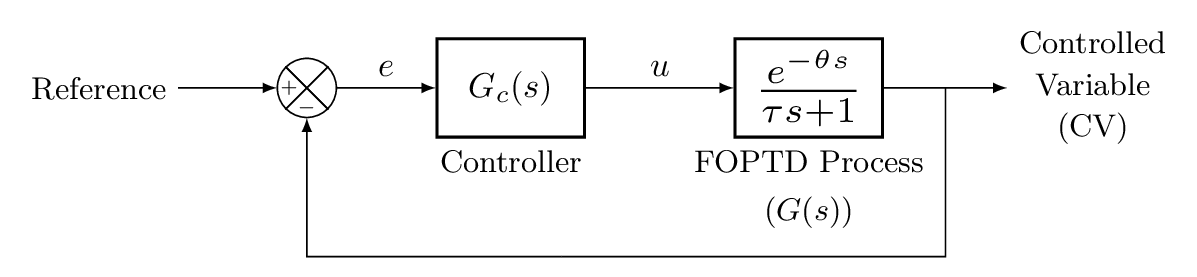}
    \caption{Block Diagram of the FOPTD system. $\theta$ and $\tau$ are the time delay and time constant of the FOPTD, respectively.}
    \label{fig:foptdblock}
\end{figure}

The block diagram for the FOPTD model under consideration is shown in Figure \ref{fig:foptdblock}, where we have assumed unity feedback of the system output. $G_c(s)$ represents the transfer function of the plant's controller, which is of the general PID form:

\begin{equation}
    \label{eq:fullpidform}
    G_c(s) = K_p + \frac{K_i}{s} + K_ds,
\end{equation}

where $K_p$, $K_i$, and $K_d$ are the proportional, integral, and derivative controller gains, respectively.

Consider the following example FOPTD process model:

\begin{equation}
\label{eq:systf}
    G(s) = \frac{e^{-0.3s}}{s+1},
\end{equation}
where the time delay, $\theta = 0.3$ second and the time constant, $\tau = 1.0$ second.

\subsection{Case 1: Time Delay Approximated using Padé Function}
For a time delay, $e^{-\theta s}$, the general 1/1 (first-order) Padé approximation for the delay is given as:

\begin{equation}
    e^{-\theta s} = \frac{1 - \frac{\theta}{2}s}{1 + \frac{\theta}{2}s},
\end{equation}

assuming $\theta \ll \tau$. Hence, with the Padé function, we write the exponential term in (\ref{eq:systf}) as:

\begin{equation}
    e^{-0.3s} = \frac{1 - \frac{0.3s}{2}}{1 + \frac{0.3s}{2}}.
\end{equation}

The FOPTD, with the delay approximated, can then be re-written as:

\begin{equation}
    \label{eq:foptdapprox}
    G(s) = \frac{e^{-0.3s}}{s+1} = \frac{1 - \frac{0.3s}{2}}{1 + \frac{0.3s}{2}},
\end{equation}
which is simplified to give:
\begin{equation}
    \label{eq:simplefoptdmod}
    G(s) = \frac{-0.3s+2}{0.3s^2 + 2.3s + 2}.
\end{equation}

From Figure \ref{fig:foptdblock} and the expression in (\ref{eq:simplefoptdmod}), with a proportional-only controller of gain $K_p = K$, we can write the closed-loop transfer function of the system as:

\begin{equation}
    \label{eq:closedlooptf}
    \frac{-0.3Ks+2K}{0.3s^2+(2.3-0.3K)s+(2+2K)}.
\end{equation}

\mypara{\textit{Controller Design for Case 1}}\hfill\\
By the R-H stability criterion, for the system in (\ref{eq:closedlooptf}) to be stable, the elements in the first column of the Routh array generated from its characteristic equation must all be positive. Table \ref{tab:routhcase1} is the Routh array corresponding to (\ref{eq:closedlooptf}).

\begin{table}[H]
    \caption{Routh Array for Case 1}
        \label{tab:routhcase1}
    \centering
    \begin{tabular}{c|cc}
       $s^2$  & 0.3 & $2+2K$ \\
       \hline
         $s^1$  & $2.3-0.3K$ & $0$\\
          $s^0$  & $2+2K$ & $0$\\
    \end{tabular}

\end{table}

\begin{figure}[H]
    \centering
    \includegraphics[trim=35pt 0 20 0,clip, width=\columnwidth]{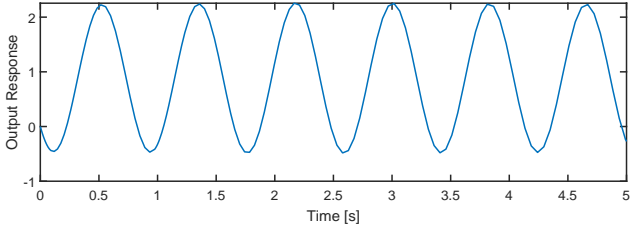}
    \caption{Response with Oscillations of Constant Amplitude for Case 1 ($K= 7.67$).}
    \label{fig:constamp}
\end{figure}

Thus, by the R-H criterion, $2.3-0.3K > 0$ and $2+2K > 0$. Hence, the range of values that will ensure the stability of the system is $-1 < K < 7.67$. To tune a PID controller using the Z-N tuning rules (Table \ref{tab:zntune}), we require two parameters - the ultimate gain of the system $k_u$, and the period of oscillation $T_u$. By definition, the ultimate gain is equal to the maximum value in the interval for system stability, that is, $k_u = 7.67$. Setting $K = k_u$ as the controller gain yields a closed-loop response with oscillations of constant amplitude, depicted in Figure \ref{fig:constamp}.

On the other hand, $T_u$ is calculated from Figure \ref{fig:constamp} as $0.8219$ second. With $k_u = 7.67$ and $T_u = 0.8219$, we compute the controller parameters presented in Table \ref{tab:contparam}. The corresponding values for $K_i$ and $K_d$ are computed using the well-known relationships:
\begin{equation*}
    K_i = \frac{K_p}{\tau_i}; \ K_d = K_p\tau_d,
\end{equation*}
with $\tau_i$ and $\tau_d$ equal to the integral and derivative times, respectively.
\begin{table}[H]
    \caption{Ziegler-Nichols Tuning Rules}
    \label{tab:zntune}
    \centering
    \begin{tabular}{cccc}
    \thickhline
     \textbf{Controller Type}    & $K_p$ & $\tau_i$ & $\tau_d$  \\
     \hline
     \textbf{P} & $0.5k_u$ & - & - \\
     \textbf{PI} & $0.45k_u$ & $0.83T_u$ & -\\
     \textbf{PID} & $0.6k_u$ & $0.5T_u$ & $0.125T_u$\\
\thickhline
    \end{tabular}
\end{table}

\begin{table}[H]
    \caption{Controller Settings for Case 1 (with $k_u = 7.67$ and $T_u = 0.8219$)}
    \label{tab:contparam}
    \centering
    \begin{tabular}{cccc}
    \thickhline 
     \textbf{Controller Type}    & $K_p$ & $\tau_i$ & $\tau_d$  \\
     \hline
     \textbf{P} & $3.83$ & - & - \\
     \textbf{PI} & $3.45$ & $0.428$ & -\\
     \textbf{PID} & $4.6$ & $0.411$ & $0.103$\\
     \thickhline 
    \end{tabular}
\end{table}

From (\ref{eq:closedlooptf}), the closed-loop transfer function of the system with the time delay approximated and with the PID settings $K_p = 4.6$, $K_i = 11.194$, and $K_d = 0.473$ is given by:

\begin{equation}
    \label{eq:evalcltf}   
    \small
    G_{CL \approx} = \frac{ -0.1418 s^3 - 0.4348 s^2 + 5.842 s + 22.39}{0.1582 s^3 + 1.865 s^2 + 7.842 s + 22.39}.
\end{equation}

\subsection{CASE 2: Time Delay Term retained in the Process Model}
\begin{figure}[H]
    \centering
    \includegraphics[trim=0pt 0pt 0pt 33pt,clip, width=\columnwidth]{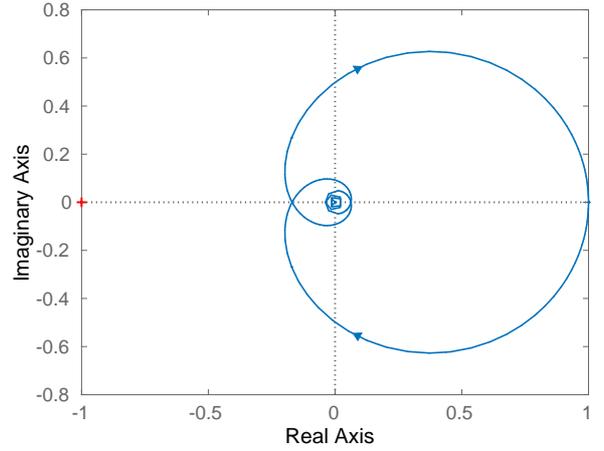}
    \caption{Nyquist plot for the FOPTD process.}
    \label{fig:nyqplot}
\end{figure}

For this case, we apply a different approach to obtain the critical gain and period of the system's response by reading off the gain margin and margin frequency from the Nyquist plot of the FOPTD process. Figure \ref{fig:nyqplot} shows the obtained Nyquist plot. We compute the critical gain, $k_u$ and period, $T_u$, from (\ref{eq:critgain}) and (\ref{eq:critperiod}), respectively.

\begin{figure}[H]
    \centering
    \includegraphics[ width=1\columnwidth]{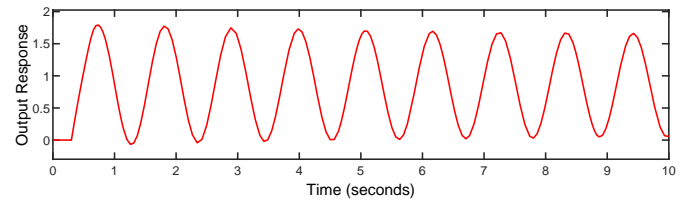}
    \caption{Response with Oscillations of Constant Amplitude for Case 2 ($K= 5.8902$).}
    \label{fig:constampnq}
\end{figure}

\setcounter{para}{0}
\mypara{\textit{Controller Design for Case 2}}\hfill\\
The critical gain is given by:
\begin{equation}
    \label{eq:critgain}
    k_u = 10^{\frac{G_m(\text{dB})}{20}},
\end{equation}
where 
\begin{equation}
    G_m(\text{dB}) = 20 \text{log} G_m(\text{mag}),
\end{equation}
while the critical period is obtained via:
\begin{equation}
 \label{eq:critperiod}
    T_u = \frac{2\pi}{\omega_c}.
\end{equation}
Here, $\omega_c$ is the margin frequency, $G_m(\text{mag})$ is the gain margin magnitude, and $G_m(dB)$ is the gain margin in decibels (dB). From the Nyquist plot, $G_m(\text{dB}) = 15.4026$ dB and $\omega_c = 5.8047$ rad/s. Applying (\ref{eq:critgain}) and (\ref{eq:critperiod}), we obtain $k_u = 5.8902$ and $T_u = 1.0824$ s. To confirm these values, we simulate the response with the gain of the proportional-only controller equal to 5.8902. As expected, a response with oscillations of constant amplitude (shown in Figure \ref{fig:constampnq}) is obtained for $k_u = 5.8902$.

Similarly, we apply the Z-N tuning rules in Table \ref{tab:zntune} in the derivation of the controller parameters shown in Table \ref{tab:case2pid}.

\begin{table}[H]
    \caption{Controller Settings for Case 2}
    \label{tab:case2pid}
    \centering
    \begin{tabular}{cccc}
    \thickhline 
     \textbf{Controller Type}    & $K_p$ & $\tau_i$ & $\tau_d$  \\
     \hline 

     \textbf{P} & $2.9451$ & - & - \\
     \textbf{PI} & $2.6506$ & $0.8984$ & -\\
     \textbf{PID} & $3.5341$ & $0.5412$ & $0.1353$\\
     \thickhline 
    \end{tabular}
\end{table}

The closed-loop transfer function of the system, with the time delay retained in the FOPTD process and the PID settings $K_p$ = 3.5341, $K_i$ = 6.5301, and $K_d$ = 0.4782, is given by

\begin{equation*}
        \label{eq:evalcltfnoapprox}
        \resizebox{.95\hsize}{!}{
    $G_{CL \not\approx} = \frac{e^{-0.3s}(0.48s^2+3.53s+6.53)}{(1+0.48e^{-0.3s})s^2+(1+3.53e^{-0.3s})s+6.53e^{-0.3s}}$
    }.
\end{equation*}

Unlike the transfer function in (\ref{eq:evalcltf}), here, the time delay component appears in the closed-loop transfer function, and there is an input and output delay in the transfer function, which is undesirable, thus motivating the need for a delay-approximation based technique.

\subsection{IMC Controller Settings for the FOPTD Process}
The methodical derivation of IMC controller settings for dynamical systems such as the FOPTD model, can be found in detail in the literature (\cite{seborgProcessDynamicsControl2010,riveraInternalModelControl1986}). Following (\cite{seborgProcessDynamicsControl2010}), by Taylor series expansion, we know that
\begin{equation}
\label{eq:tayapr}
    e^{-\theta s} \approxeq 1-\theta s,
\end{equation}

for $\theta$ small enough. Thus, from (\ref{eq:systf}), the transfer function of the (IMC-tuned) feedback controller can then be written as:
\begin{subequations}
    \begin{align}
    G_c &= \frac{s+1}{(\tau_c + \theta)s}\\
        \label{eq:gcimc}
        &= \frac{s+1}{1.8s} \equiv 0.555 + \frac{0.555}{s},
\end{align}
\end{subequations}

where we have chosen the desired closed-loop time constant ($\tau_c$) to be equal to $1.5$. From (\ref{eq:fullpidform}), it is clear that (\ref{eq:gcimc}) is equivalent to a parallel PI-controller with $K_p = K_i$ = 0.555. We derive the controller settings for the SIMC-tuned controller in the next subsection.

\subsection{SIMC (Skogestad-IMC) Controller Parameters for the FOPTD Process}
Generally, the SIMC PI-settings for an FOPTD process of the form 
\begin{equation}
\label{eq:gceq}
    G(s) = \frac{k}{\tau_1 s + 1}\cdot e^{-\theta s},
\end{equation}

as proposed by (\cite{grimholtOptimalPIControlVerification2012}), are given as:
\begin{equation}
\label{eq:kceq}
    K_c = \frac{1}{k}\cdot\frac{\tau_1}{\tau_c + \theta},
\end{equation}
where 
\begin{equation}
    \label{eq:tieq}
    \tau_i = \text{min}\{\tau_1, 4(\tau_c + \theta)\},
\end{equation}

and $\tau_c$ retains its previous definition. In what follows, we shall consider two cases with different values of the desired closed-loop time constant: the value for tight tuning and that for smoother tuning, as will be indicated in the following subsections.
\setcounter{para}{0}
\mypara{\textit{For Tight Tuning ($\tau_c = \theta$)}}\hfill\\
For this case, using (\ref{eq:kceq}) and (\ref{eq:tieq}), with $\tau_1 = k = 1$ and $\theta = 0.3$, we obtain the values of $K_c$ and $\tau_i$ as 1.67 and 1 s, respectively. Hence, the proportional and integral gains of the PI controller become 1.67, i.e. $K_p = K_i = 1.67$. The parallel form of the PI controller thus becomes:
\begin{equation}
    G_c(s) = 1.67\bigg(1+\frac{1}{s}\bigg).
\end{equation}

\mypara{\textit{For Smoother Tuning ($\tau_c = 1.5\theta$)}}\hfill\\
For smoother tuning, we compute the values of $K_c$ and $\tau_i$ as 1.33 and 1 s, respectively, also from (\ref{eq:kceq}) and (\ref{eq:tieq}). $K_p$ and $K_i$ are thus, both equal to 1.33 in this case. The resulting PI controller is therefore:
\begin{equation}
    G_c(s) = 1.33\bigg(1+\frac{1}{s}\bigg).
\end{equation}

\section{Numerical Simulations}
\label{sec:simex}
For software simulations, we employ $\text{MATLAB/Simulink}$, with the PID controllers for both instances discussed, together with their respective process models, nested in two independent subsystems. 

\subsection{Simulation Results for the two Controller Cases}

\begin{figure}[H]
    \centering
    \includegraphics[scale=0.6,trim=30pt 0pt 15pt 20pt,clip]{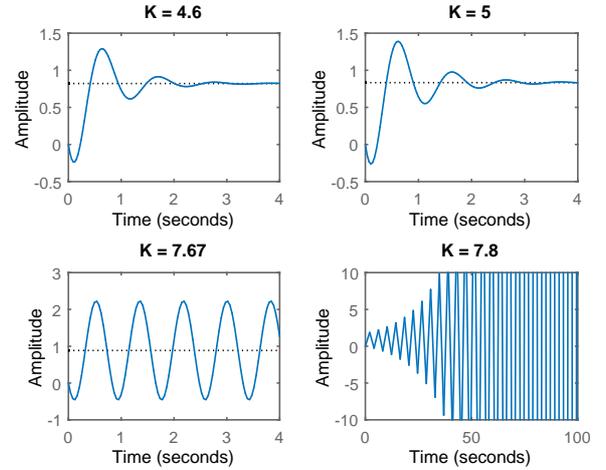}
    \caption{Closed-Loop Step Response of the FOPTD model for different values of controller gain.}
    \label{fig:piddiffgain}
\end{figure}

For Case 1, the closed-loop response of the FOPTD model for different values of controller gain is shown in Figure \ref{fig:piddiffgain}.

Considering the interval for stability for the system ($-1<K<7.67$), from the graphs in Figure \ref{fig:piddiffgain}, we can see that the response grows unbounded and is unstable beyond $K=7.67$. On the other hand, the response is stable and hardly oscillatory for $K = 4.6$ and $K = 5$, which validates the accuracy of the interval. As already discussed, $K=7.67$ leads to a response with oscillations of constant amplitude. Another interesting observation is that the system is minimum-phase, for the values of $K$ within the interval of stability. A system is minimum-phase if all of its poles and zeros lie in the left-half $s$-plane (\cite{ogataModernControlEngineering2015}). The poles of the system for $K = 4.6$ and $K=5$ are given in Table \ref{tab:poles}.

\begin{table}[htb]
    \caption{Poles of the Closed-loop transfer function for $K$ = 4.6 and 5}
    \label{tab:poles}
    \centering
    \begin{tabular}{cccc}
    \thickhline 
    \multirow{2}{*}{\textbf{$K$}} & \multirow{2}{*}{\textbf{Poles}} & \multirow{2}{*}{\textbf{Location of}} & \multirow{2}{*}{\textbf{Location of}}\\
    & &\textbf{$1^{\text{st}}$ pole} &\textbf{ $2^{\text{nd}}$ pole}\\
    \hline
       \multirow{2}{*}{4.6}  &  $  -1.5333 + j5.9146$, & \multirow{2}{*}{Left-half $s$-plane} & \multirow{2}{*}{Left-half $s$-plane}\\
       & $-1.5333 - j5.9146$ & &\\
       \hline
      \multirow{2}{*}{5}  &  $-1.3333 + j6.1824$,& \multirow{2}{*}{Left-half $s$-plane} & \multirow{2}{*}{Left-half $s$-plane}\\
      & $-1.3333 - j6.1824$ & &\\
    \thickhline 
    \end{tabular}
\end{table}

\begin{figure}[H]
    \centering
    \includegraphics[trim=30pt 10pt 30pt 10pt, clip, width=\columnwidth]{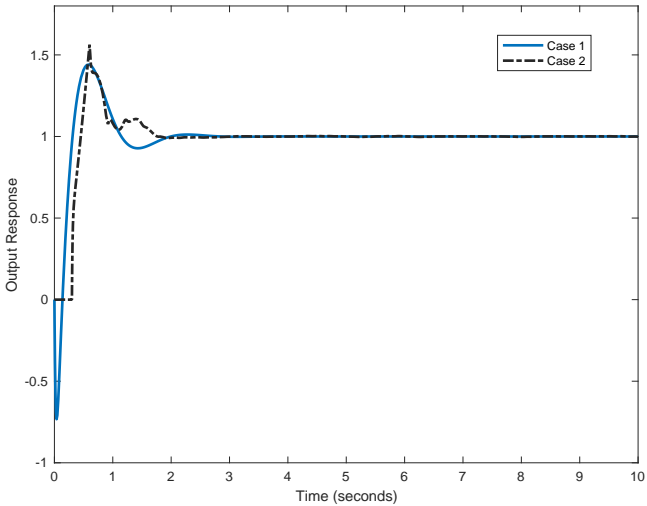}
    \caption{Comparison of step responses of the FOPTD system for Case 1 and Case 2.}
    \label{fig:twocases}
\end{figure}

With a unit step input, applied as a reference signal at $t=0$ s, the response for the two case studies is given in Figure \ref{fig:twocases}. Unlike the delay-approximated case, the output with the time delay retained in the model (Case 2) is chaotic, with recurrent jumps at different time intervals. This response is typical of time-delayed systems, and these jumps signify discontinuities in the system output (\cite{StepResponsePlot,shampineDelaydifferentialalgebraicEquationsControl2006}).

For a quantitative evaluation, we compare performance metrics of the step responses for the two controller cases in Table \ref{tab:perfcomp},. $e_{ss}$ is the steady-state error.

\begin{table}[H]

    \centering
    \caption{Performance Characteristics of the FOPTD system for the two cases}
    \label{tab:perfcomp}
    \resizebox{\columnwidth}{!}{%
    \begin{tabular}{cccccc}
   \thickhline 
    \multirow{2}{*}{\textbf{Case}} & {\textbf{Settling time},} & {\textbf{Rise time},} & {\textbf{Peak}} & {\textbf{\%}} & {\textbf{$e_{ss}$}} \\
    & \textbf{$t_s$ [s]} & \textbf{$t_r$} [s] &\textbf{Amplitude} & \textbf{Overshoot} & \\
    \hline
    1 & 1.83 & 0.13 & 1.44 & 44.1 & 0\\
    \hline
    2 & 1.70 & 0.12 & 1.56 & 56.4 & 0\\
  \thickhline 
    \end{tabular}%
    }
\end{table}

    From Table \ref{tab:perfcomp}, we can infer that the system response, with the time delay approximated, has a comparably fast response as the case with the time delay retained, but also with more accurate tracking performance and less overshoot.

\subsection{Simulation Results with IMC and SIMC controller settings}

\begin{figure}[H]
    \centering
    \includegraphics[width=1\columnwidth]{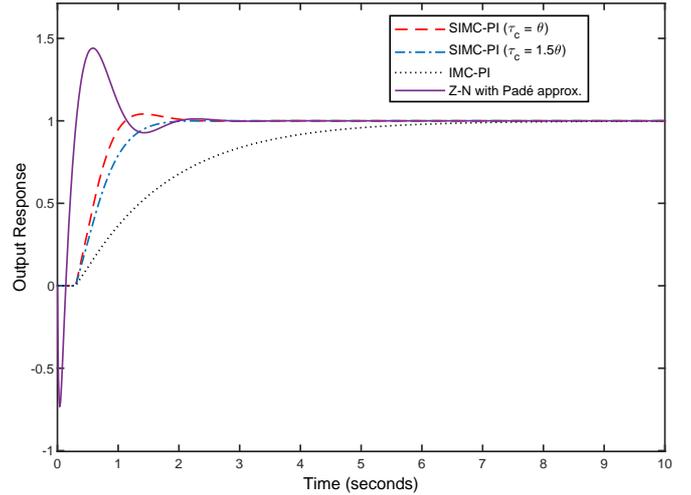}
    \caption{Comparative step response of the FOPTD process with SIMC-PI controller settings, IMC-PI settings, and PID controller parameters (Z-N tuned, with Padé Approximation).}
    \label{fig:imcvssimc}
\end{figure}

Figure \ref{fig:imcvssimc} presents a comparative response of the SIMC-PI, IMC-PI and Padé-approximated PID controller settings. Table \ref{tab:perfchar} summarizes the performance characteristics of the simulated controllers, from where it can be seen that with Padé approximation, the PID controller leads to superior tracking performance. Damping ratios of 0.473 (twice) and 1 are obtained in the case with Padé approximation, which suggests an acceptably-damped system response (\cite{roskillyChapterFiveClosedLoop2015}).

\begin{table*}
    \caption{Performance Characteristics of the Simulated Controllers}
    \label{tab:perfchar}
    \centering
    \begin{tabular}{ccccc}
    \thickhline 
    \multirow{3}{*}{\textbf{Parameter}} & {\textbf{Response with Z-N tuning}} & {\textbf{Response with}} & {\textbf{Response with}} & {\textbf{Response with}}\\
    &\textbf{(with Padé Approximation)} & \textbf{IMC-PI settings}& \textbf{SIMC-PI settings} & \textbf{SIMC-PI settings}\\
    &&& ($\tau_c = \theta$) & ($\tau_c = 1.5\theta$)\\
    \hline
    $K_p$ & 4.600 & 0.555 & 1.67 & 1.33  \\
    $K_i$ & 11.194 & 0.555 & 1.67 & 1.33 \\
    $K_d$ & 0.473 & - & -& -\\
    $t_s$ [s] & 1.83 & 5.99 & 1.82 & 1.64 \\
    $t_r$ [s] & 0.14 & 3.22 & 0.57 & 0.85\\
    \% OV & 44.08 & 0.00 & 4.12 & 0.07 \\
    {Peak amp.} & 1.44 & 0.999 & 1.04 & 1.00 \\
    $e_{ss}$ & 0.00 & 0.00 & 0.00 & 0.00\\
   \thickhline 
    \end{tabular}

\end{table*}

\section{Conclusion \& Future Work}
\label{sec:conc}
We presented results on controller tuning for FOPTD processes, based on delay approximation, where we showed that, with the time delay retained in the FOPTD model, an undesirable chaotic response with discontinuities in the plant output is observed. In contrast, the IMC and SIMC tuning rules give good setpoint tracking and excellent stability, as the corresponding responses are overdamped and stable. They also yield a non-minimum phase plant. In contrast, a minimum phase system is obtained with the delay in the FOPTD model approximated using the Padé function and its PID controller tuned with Z-N tuning rules.

A considerable level of overshoot is observed in the response with Padé approximation of the delay, but with a faster settling and rise time than the response with the IMC-PI controller settings. At the same time, the IMC-PI and SIMC-PI settings produce a more stable response with very minimal overshoot. In particular, the output with the SIMC-PI controller settings (for $\tau_c = 1.5\theta$) has almost zero overshoot and good setpoint tracking. Furthermore, the results in Table \ref{tab:perfchar} prove that for the FOPTD system, delay approximation via the Padé function eliminates the undesirable effects of the time delay and guarantees a faster tracking performance superior to the controllers tuned with the IMC and SIMC (tight) tuning approaches. Nevertheless, there is still some overshoot to the delay-approximated controller's performance. Hence, the need to further optimize the controller to produce a more desirable plant behavior, namely robust tracking with zero overshoot. As Padé approximation is only valid for delays much smaller than the plant's time constant, questions about the robustness of our proposed control scheme to variable delay will arise. Future research will focus on these considerations.

\bibliography{ifacconf}

\end{document}